\documentclass[pre,amsmath,aps,superscriptaddress,twocolumn]{revtex4}
\usepackage{graphicx,xspace}
\usepackage{float}
\usepackage{amsmath}
\usepackage{amssymb}
\usepackage{color}
\usepackage[normalem]{ulem}
\input{epsf}

\begin{document}

\title{The mechanisms of spatial and temporal earthquake clustering}

\author{E. A. Jagla}
\affiliation{Centro At\'omico Bariloche, Comisi\'on Nacional de Energ\'{\i}a At\'omica, 8400 Bariloche, Argentina}
\thanks{E-mail: jagla@cab.cnea.gov.ar}
\author{A. B. Kolton}
\affiliation{Centro At\'omico Bariloche, Comisi\'on Nacional de Energ\'{\i}a At\'omica, 8400 Bariloche, Argentina}

\begin{abstract}
The number of earthquakes as a function of magnitude decays as a power law. This trend is usually justified using spring-block models, where slips with the appropriate global statistics have been numerically observed. However, prominent spatial and temporal clustering features of earthquakes are not reproduced by this kind of modeling. We show that when a spring-block model is complemented with a mechanism allowing for structural relaxation, realistic earthquake patterns
are obtained. The proposed model does not need to include a phenomenological velocity weakening friction law, 
as traditional spring-block models do, since this behavior is effectively induced by the relaxational mechanism as well. 
In this way, the model provides also a simple microscopic basis for the widely used phenomenological rate-and-state 
equations of rock friction.
\end{abstract}

\maketitle

\section {Introduction}

The distribution of earthquakes in nature follows non-trivial patterns, some of which are captured by well known 
empirical laws.
The Gutenberg-Richter (GR) law \cite{scholz,gr}
states that the number of earthquakes as a function of magnitude $N(M)$ scales as $N(M)\sim 10^{-bM}$. The exponent $b$ is very nearly 1. The Omori law refers to temporal correlations between earthquakes, in particular
to aftershocks, namely the temporal clustering of earthquakes following a large one, usually called the main shock.
The Omori law of aftershocks\cite{omori} states that the number of aftershocks per unit of time decays as 
$\sim (t+c)^{-p}$ with the time $t$ from the main shock. The exponent $p$ is typically very close to 1, and $c$ is a 
time constant of the order between minutes and hours. Aftershocks occur mainly in the spatial region where the rupture of the main shock took place. 

The GR law has been shown to be compatible with a state of (at least partial) critical organization of the system\cite{bak,bak2,langer} that is understandable in terms of spring-block models, when an appropriate velocity weakening friction law (i.e., a friction force decreasing with the relative velocity of the sliding elements) is assumed to hold. This kind of modeling was pioneered by Burridge and Knopoff\cite{bk} (BK), and was extended along different directions afterwards, particularly in the works on self-organized-criticality of the eighties \cite{bak,bak2}.
The BK model reproduces the global statistical behavior implied by the GR law, but it fails to account for the existence of spatial and temporal correlations observed in actual seismicity. 
On searching for the origin of the aftershock phenomenon, Dieterich\cite{dieterich94} followed by others 
\cite{marcellini,ziv,moreno,shaw} have shown that an
analysis based on rate-and-state equations\cite{dr1,dr2} is able to justify the appearance of aftershocks following the Omori decay.
On this perspective, it is puzzling that the use of a BK model with a rate-and-state friction law does not produce realistic aftershocks\cite{kawamura}.
Although aftershocks usually are responsible for less than about 5 $\%$ of the total released seismic moment, 
the finding that the GR law is also obeyed within individual aftershock sequences strongly 
suggests that consistent and compatible explanations for GR and Omori laws should exist. 
We may thus say that at present,
there is not a single, unified picture of the physics behind some 
of the most robust features of seismicity, namely GR and Omori laws. In addition, the use 
of rate-and-state equations, although widely
supported by experimental results, remains essentially a crude phenomenological approach.

We show here that when a spring-block model without any a priori velocity 
dependent frictional force is 
complemented with an appropriate relaxational term as discussed below, it 
produces: 1) earthquake patterns and in particular aftershock sequences quantitatively comparable with real ones,
2) a velocity weakening friction law, and in general, agreement with the predictions of the
rate-and-state equations, and 3) a power law decay of number of earthquakes with magnitude
compatible with the GR law, with an exponent $b$ that compares well with actual values.

\section{Model in the absence of relaxation}

Our modeling is based on the original BK model\cite{bk}, with the important difference that the
friction law between the blocks and the substrate is not a priori assumed to have any particular 
form such as the velocity-weakening form commonly used. A velocity dependent friction law emerges naturally at 
large scales from the characteristic collective dynamics of elastic manifolds in 
random media\cite{brazovskii}. In this context, 
an elastic interface (which corresponds to the blocks joined by springs in the BK model - we already 
describe the two dimensional case, more appropriate to real faults) is driven through a disordered 
potential energy landscape (the `substrate') that models the random nature of asperities. The 
velocity-dependent frictional force between the blocks and the substrate of 
the BK model is therefore replaced by a disordered potential energy 
landscape that is chosen randomly and uncorrelated for each point block  
of the discretized interface. 
In concrete, our model is described by the overdamped equation
\begin{equation}
\lambda \frac{\partial u_{i,j}}{\partial t}=k_0\nabla^2 u_{i,j}+f_{i,j}+k_1(X_0(t)-u_{i,j})
\label{lambda}
\end{equation}
where $u_{i,j}$ is a continuous variable representing the displacement of the block labeled 
by the indices $(i,j)$ in a two dimensional grid, 
$X_0(t)$ is the driving variable (usually $X_0(t)=Vt$, we will refer also to $X_0$ and to $k_1(X_0(t)-u_{i,j})$
as the strain and local stress in the system, respectively), $\nabla^2$ is the discrete Laplacian operator, 
and $f_{i,j}$ is the pinning force at each block, which is assumed to be 
short-range correlated along the direction of the block displacements. 
For numerical convenience this spatially random force is chosen in the following way:
a random position $u^0_{i,j}$ is selected for each $i,j$, then the force is $f_{i,j}=k(u_{i,j}-u^0_{i,j})$.
When, upon the dynamical evolution, $f_{i,j}$ reaches some threshold value $f^{th}_{i,j}$ (that is chosen
randomly distributed between $1+\kappa$ and $1-\kappa$), the corresponding $u^0_{i,j}$ is given the 
value $u_{i,j}+ \delta$, where $\delta$ is randomly chosen between $-1$ and 1. Also, a new threshold 
is assigned to site $i,j$. Similar results are obtained by using standard, 
though computationally more demanding, short-ranged correlated smooth 
pinning forces such as the ones used in Refs. \cite{lacombe,alberto}.

We use periodic boundary conditions, and from now on, we set the values $k_0=0.1$, $k=1$, $\kappa=0.8$. Taking also the numerical lattice constant as unity, this renders our time, distance, and forces, dimensionless. We have tried other parameter sets, finding no qualitatively new results. The value of $\lambda$ in Eq. (\ref{lambda}) fixes the time scale necessary for the surface to adapt to the conditions dictated by $u^0$ and $X_0$. We work in the case $\lambda\rightarrow 0$, i.e., 
we give the surface time to relax to what we call a meta-stable configuration, defined by equating the right hand side of Eq. (\ref{lambda}) to zero, for fixed values of $X_0$ and $u^0_{i,j}$. Note in particular that this also means that the duration of individual earthquakes is zero in our implementation.

\begin{figure}
\centerline{\includegraphics[width=.5\textwidth]{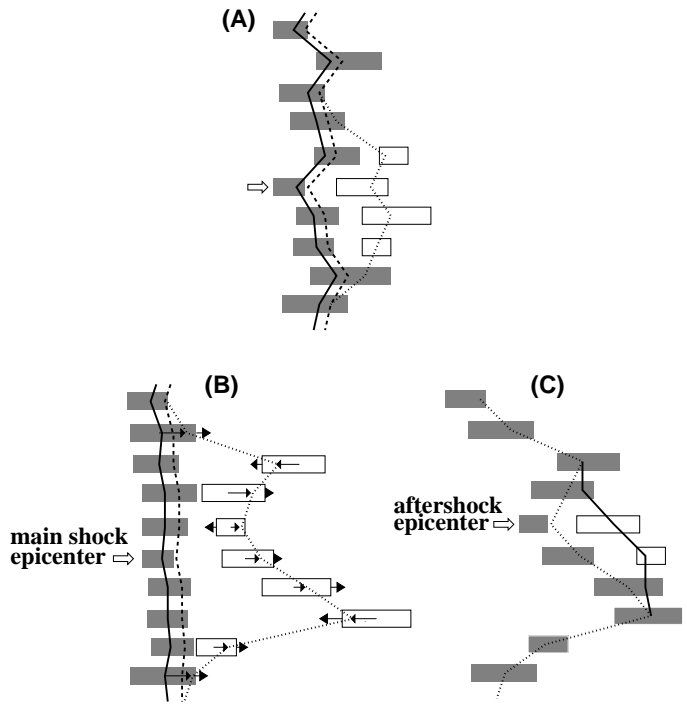}}
\caption{One dimensional sketch of main processes that occur in our model. (A) In the absence of relaxation, the
interface (vertical line, with coordinates $u_{i,j}$) is driven to the right by an external force (not shown) through a set of randomly placed pinning centers, represented by the gray rectangles. The mid point of the rectangles have coordinates $u^0_{i,j}$, and the horizontal length is the range in which pinning is effective. Different lengths indicate different threshold values $f^{th}_{i,j}$. Upon driving, the system passes from the configuration indicated by the continuous line to the dashed line, and an event is triggered at the site indicated by the arrow, where the maximum local pinning force is overpassed. The system goes through a cascade process (not fully indicated) onto a new meta-stable configuration (dotted line) in which some pinning centers have been refreshed (outlined rectangles). In (B) structural relaxation is acting. A quite relaxed (and therefore more coherently pinned) initial configuration (continuous line) is driven until
a main shock occurs, at the configuration corresponding to the dashed line. After the system has reached a meta-stable configuration (dotted line and outlined rectangles) relaxation 
continuous to act modifying the position of the pinning centers. The arrows at the center of the rectangles indicate the local value of $(u-u^0)$. The arrows just outside the rectangles indicate the values of  $du^0/dt$ according to Eq. (\ref{eq2}), that produce a drift in the position of the pinning centers. This drift may cause a further instability as can be seen in (C). The process in (C) is an aftershock to the main shock in (B).
}
\label{f0}
\end{figure}

Abrupt rearrangements occur in the system whenever at some particular position $i,j$  the force from the pinning potential of the substrate is not able to sustain any more the surface pinned to it (see an example
of this situation in Fig. \ref{f0}A).
In this situation the local rearrangement of the surface can trigger instability events in neighbor sites, and the process continues until the surface finds a new globally stable configuration. 
The full sequence of all rearrangements triggered by an initial instability is what we call an event,
or an earthquake, being the position of the triggering instability its epicenter. We measure the seismic moment 
$m_0$ of events as $m_0 = \sum_{i,j} \Delta_{i,j}$ 
where $\Delta_{i,j}$ is the displacement caused by the event at position $i,j$. 
In order to compare with real earthquakes, the magnitude $M$ of an event is defined from the
seismic moment as $M = 2/3\log_{10}m_0$.

\subsection{Results}

\begin{figure}
\centerline{\includegraphics[width=.4\textwidth]{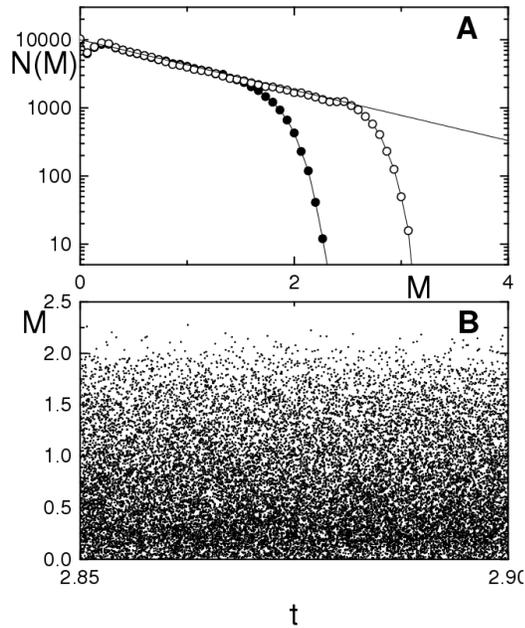}}
\caption{Results without relaxation. ({\bf A}) Magnitude histogram for systems of 512x512 sites, with $k_1=0.01$ (full symbols) and $k_1=0.001$ (open symbols). Continuous line has a slope $b=0.4$. ({\bf B}) Magnitude-time plot
for a system of size 256x256, $k_1=0.01$.
}
\label{S1}
\end{figure}

Our model in the absence of relaxation has been widely studied, and is known 
to have a well defined size distribution 
of events\cite{lacombe,alberto,zapperi,pierre}
(note that the usual definition of the decay exponent $\tau$ is given as a function of our $b$ 
as $\tau=2b/3+1$). In Fig. \ref{S1}A we show this distribution.
We observe a power law with exponent $b\simeq 0.4$, which is consistent with 
the expected results for two dimensional elastic interfaces both from scaling arguments\cite{zapperi}
 and from recent analytical and numerical calculations\cite{pierre}. This value 
is however well different from the $b\simeq 1$ observed for actual earthquakes.  
An exponential cut-off for large event size exists due to confinement. 
This cut-off is controlled by the rigidity $k_1$ of the 
driving spring and occurs when the spatial extent of the events
in the direction of the displacements is of order $k_1^{-\zeta/2}$, 
with $\zeta$ the interface roughness exponent at low velocities. 
The crossover to the exponential behavior thus moves to larger 
magnitudes as $k_1$ is decreased~\cite{lacombe,alberto,zapperi,pierre}.

In Fig. \ref{S1}B we show a time-magnitude plot of all events occurring in
a particular time interval. It is apparent that no obvious temporal correlations occur in this case, and more 
quantitative observations confirm this fact. Spatial correlations are not observed neither. This is 
qualitatively similar to what has been obtained for the original BK model\cite{bk}.

\section{Model in the presence of relaxation}

\begin{figure}
\centerline{\includegraphics[width=.4\textwidth]{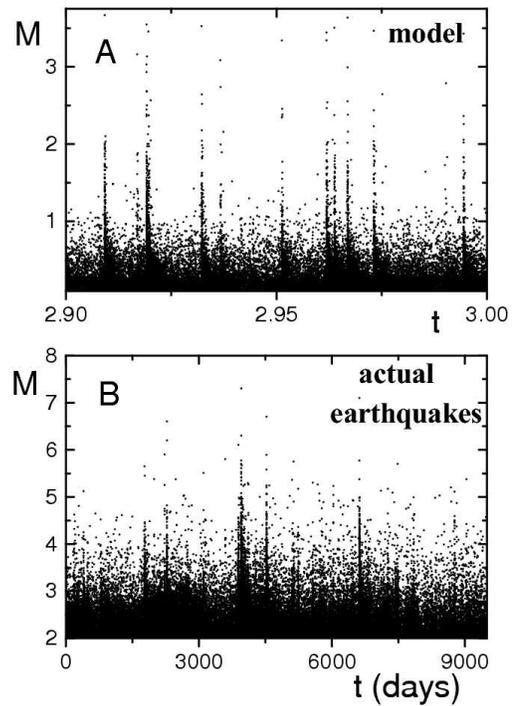}}
\caption{Results for the model with relaxation as compared to earthquakes in California\cite{California}. ({\bf A})Magnitude-time plot for a 512x512 system in the presence of relaxation ($k_1=0.01$, $R/V=500$).  ({\bf B}) Actual earthquakes in the California area.
}
\label{512r500}
\end{figure}

So far, in the present form, the model does not give any clue on the reason for earthquake clustering, or the origin of a 
velocity weakening friction law. However, the inclusion of a simple additional ingredient changes this 
scenario drastically. This ingredient 
turns out to be what we have called {\it structural relaxation} \cite{jagla}. The primary physical justification of its inclusion is the following. It is known that in solid friction the friction coefficient at rest increases with the time the surfaces have been in contact \cite{marone,persson}. This is telling us about the existence of a temporal dependent mechanism that makes the 
sliding surfaces get more attached or pinned to each other when they
remain in contact for a longer period of time. In the present model, a rather simple way to include such an effect is to consider, in addition to the random pinning force that the substrate performs on the sliding surface, the reaction that the surface performs onto the substrate. If we give the substrate the possibility to react to this force, the system will gain 
pinning energy by making the substrate more correlated, so to pin better the 
correlated interface structure. This is in general a slow process, 
and the longer the surface remains in contact with the substrate, the stronger the join. This process of attachment 
is however stopped and restarted when a slip event occurs, since the values of the disordered potential 
refreshen, becoming uncorrelated again. We will show in the following that this simple mechanism 
is enough to explain, in particular, the appearance of a robust sequence of aftershocks, and the occurrence of 
a velocity weakening friction law. Our structural relaxation mechanism 
is what in other contexts is called the ageing of the material.

\begin{figure}[t]
\centerline{\includegraphics[width=.4\textwidth]{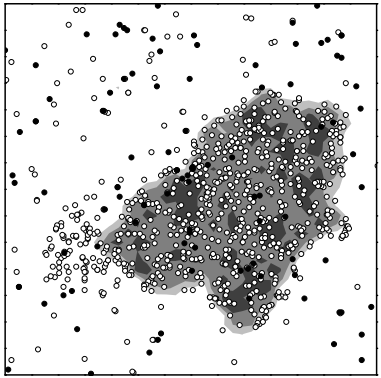}}
\caption{Spatial distribution of aftershocks in the simulations. The slip surface of a large event (shadowing proportional to the local slip) and the before- (full) and after-events (open symbols) of magnitude larger than 0
occurring in a symmetric time interval $\delta t=0.0022$  around the main event are shown. The increase in seismicity after the main shock is clearly observable, as well as the localization of aftershocks mainly at and near the slip surface of the main shock. The figure
depicts a portion of size 350x350 of a system of 512x512.
}
\label{locate}
\end{figure}

The modification to the model is as follows. We allow the values of $u^0_{i,j}$ to relax in time according to 
\begin{equation}
\frac{\partial u^0_{i,j}}{\partial t}=-R \nabla^2 f_{i,j}= Rk \nabla^2 (u^0_{i,j}-u_{i,j}).
\label{eq2}
\end{equation}
This conserved dynamics for the shift of the disorder potential at different block points is a generic way 
to introduce the back effect of the surface on the substrate (the actualization of the $u^0$'s when a slip event occurs
is made as before). The coefficient $R$ is a measure of the intensity or rate of relaxation, and can 
thus be related to experimental relaxation times. Equation 
(\ref {eq2}) generates a tendency for the local forces $f_{i,j}$ to become uniform across the system, 
generating a stronger contact between surface and substrate. This relaxational effect competes with the 
driving, which forces the movement of the surface onto the substrate at a fixed average velocity $V$. 
The relevant parameter that measures the competition between the two effects is the ratio $R/V$.


The mechanism by which earthquake clustering occurs can be summarized as follows (see Fig. \ref{f0}). If a particular region 
of the sample
has not experienced a large event in a rather long period of time, the structural relaxation has made this region stronger (Fig. \ref{f0}(B)).
When an event occurs (driven by the overall displacement between surface and substrate) the contacts refreshen and large variations in the local forces remain. Relaxation continues to act, trying to uniformize the local forces. In this process, particular points that were originally stable immediately after the main shock, may destabilize and generate a new event (Fig. \ref{f0}(C)). Note that in this description it is obvious that aftershocks will occur at, or near, the rupture region of the main shock. It is also worth noting here 
that aftershocks are triggered by the inner dynamics of the system, and that most aftershocks occur also if we 
stop the driving of the system after the main shock.
The seemingly contradictory fact that aftershocks (which must be triggered by an initial instability) are originated in a relaxation mechanism is understood when one realizes that due to the disorder,
relaxation according to Eq. (\ref{eq2}) may produce local increases in the forces $f_{i,j}$, and this can trigger aftershocks if a local threshold is overcome. Note in this respect the opposite direction between $(u-u^0)$ and $du^0/dt$ at the aftershock epicenter, in Fig. \ref{f0}C.

\subsection{Results and comparison with an actual earthquake sequence}

\begin{figure}
\centerline{\includegraphics[width=.4\textwidth]{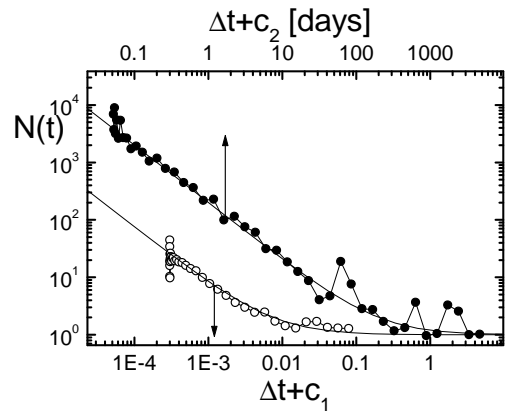}}[t]
\caption{The Omori law. Histogram of the number of events after main shocks in the simulation (open symbols), averaged over 120 main shocks, and the
histogram of aftershocks for events in the California area (full symbols), averaged over 7 events of magnitude $M > 6.0$ in the time period considered. $\Delta t$ is the time since main shock.  Curves have been vertically rescaled, setting the value 1 for large $\Delta t$. Continuous lines are fittings to Omori law with $p=1$. The time shifts are $c_1=3\times 10^{-4}$, $c_2=0.05$ days.
}
\label{S2}
\end{figure}

\begin{figure}[t]
\centerline{\includegraphics[width=.4\textwidth]{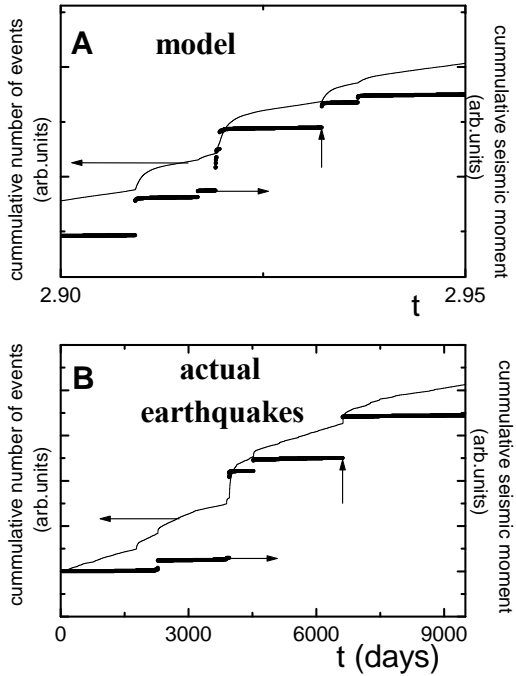}}
\caption{Cumulative number of events and cumulative seismic moment for the sequences presented in Fig. \ref{512r500}. 
}
\label{S3}
\end{figure}

In Fig. \ref{512r500}A we show a magnitude-time plot of events in a simulation of a system of 512x512 sites, in the presence of relaxation ($R/V=500$).  For comparison, the same plot for the earthquakes in the California area \cite{California} is presented as Fig. \ref{512r500}B.
The visual similarity is striking. In both graphs, the very large activity immediately after large events, i.e., the existence of aftershocks, is apparent.
It is worth noting here that a minimum value around $R/V \sim 100$ of relaxation is necessary in order to observe aftershocks in the model, so  the structural relaxation is the crucial ingredient behind these particular effects.

The spatial location of aftershocks are strongly correlated with the slip surface of the main event. In Fig. \ref{locate}
we show the region that has slip in a large event in the simulations, together with the epicenters of all events occurring in a symmetric time interval around the main shock. Events before and after the main shock are shown in a separate way.
We see that there is a rather uniform spatial distribution of before-events, whereas once the main shock has occurred, 
aftershocks occur at and near the region in which the main slip occurred.

In Fig. \ref{S2} we plot the histogram with the number of events after a main shock as a function of time.  In the simulations, we average over 120 large events in a single simulation of size 512x512. The continuous lines correspond to  Omori laws of the form 
$N(t)=A/(t-t_0+c)^{p} +N_0$, where $t_0$ is the time of the main shock and $N_0$ is the value of background seismicity. For reference, we also plot the number of aftershocks
of the seven earthquake with $M>6.0$ in the California area in the considered time period. Both cases are well fitted by an Omori law with $p\sim 1$.

\begin{figure}[t]
\centerline{\includegraphics[width=.4\textwidth]{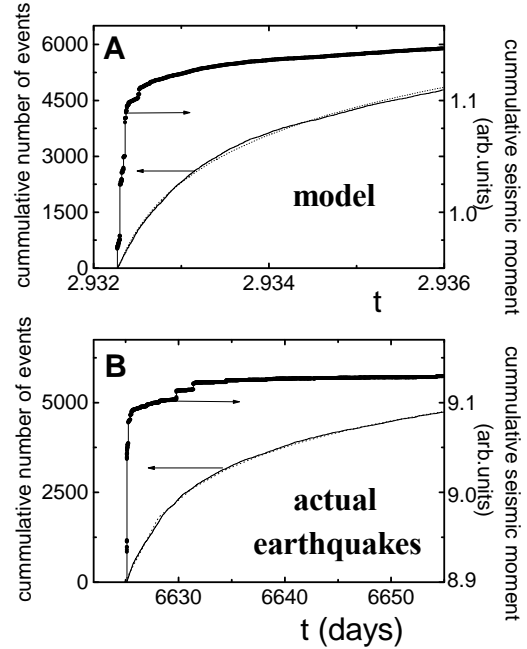}}
\caption{Detail to Fig. \ref{S3}. Cumulative number of events and cumulative seismic moment (taken as zero just before the main shock), following the events indicated by vertical arrows in Fig. \ref{S3}. In both cases, dotted lines are fitting to the (cumulative) Omori law with $p=1$.
}
\label{S4}
\end{figure}

Figure \ref{S3} shows plots of cumulated number of events and seismic moment corresponding to the sequences presented in Fig.\ref{512r500}, and Fig. \ref{S4} is a detail after the events indicated by vertical arrows in Fig. \ref{S3}.
The cumulative number of events is fitted in both cases with a cumulative Omori law, with $p=1$ with very good agreement. The evolution of the accumulated seismic moment is also qualitatively similar in both cases, with the main shock accounting for most of the released seismic moment of the whole sequence.

\begin{figure}[t]
\centerline{\includegraphics[width=.4\textwidth]{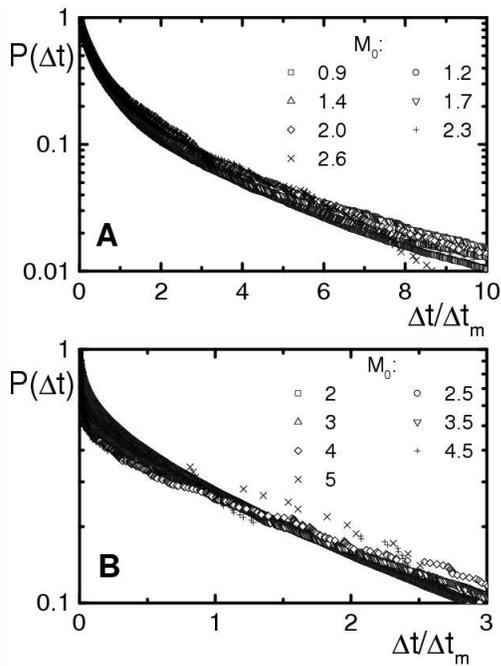}}
\caption{Time intervals distribution. Distribution of the times (normalized by the mean time $\Delta t_m$) between events of magnitude larger than $M_0$ for our data ({\bf A}) and for earthquakes in California ({\bf B}). Independence of $M_0$, and a rather exponential distribution with an excess due to aftershocks for small $\Delta t$ is observed in both cases.
}
\label{S5}
\end{figure}

Finally, in Fig. \ref{S5} we present an analysis of the time intervals $\Delta t$ between successive events of magnitude larger than some
defined threshold $M_0$. The main characteristics observed for the real sequence, that are reproduced by our model 
are the following. The curves are roughly independent of the threshold value $M_0$ chosen, and the global behavior 
represents almost an exponential decay with $\Delta t$, but with a reproducible excess of 
events at low $\Delta t$. This excess is accounted for by the aftershocks.

\section{Averaged frictional properties}

\begin{figure}[t]
\centerline{\includegraphics[width=.4\textwidth]{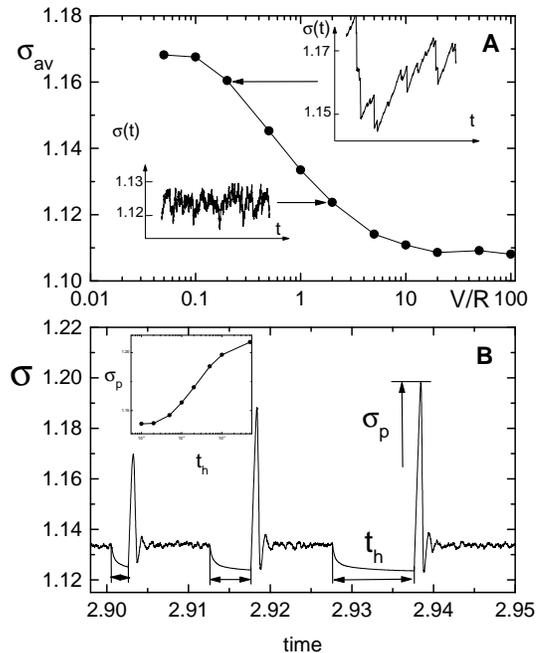}}
\caption{Global frictional properties of the model. ({\bf A}) Mean stress in a system of 256x256 as a function of relative velocity. A detail of the temporal dependence of stress is given for two points, emphasizing the larger fluctuations that appear when relative velocity is lower. ({\bf B}) Time evolution of stress in a system in which velocity is changed from $V/R=0$ in the hold periods (indicated by arrows), to $V/R=1$ in the rest of time (results shown correspond to an average over ten realizations). Inset: The value of the stress  peak as a function of the hold time.
}
\label{ss}
\end{figure}

The second set of results that we present corresponds to the stress-strain relation of the model for different driving protocols. First of all we recall that the model without relaxation shows a stress $\sigma$ that is independent of the strain rate, since the internal time scale of the model is very rapid compared to the driving. In Eq. (\ref{lambda}) this means that $\lambda/V\to 0$.
The inclusion of relaxation introduces a new time scale (set by the parameter $R$ in Eq. (\ref{eq2})) and now the
average stress in the system depends on the ratio $R/V$. When $R/V$ is small, the effect of relaxation is negligible, and the stress will be similar to that in the absence of relaxation. However, if $R/V$ is high enough, relaxation will act by effectively correlating the pinning potential in a larger spatial region. The size of this region increases with $R/V$.
A spatially more correlated pinning potential produces, in turn, a larger average stress in the system. We conclude that 
the larger is $R/V$ the larger is the average stress. In other words, the model will display velocity weakening.
In Fig. \ref{ss}A we show a plot of the stress in the system as a function of strain rate where 
this weakening is clearly observed. For large strain rates the stress converges to the value 
corresponding to no relaxation, whereas
the behavior for very small strain shows a saturation at a larger value.  The transition between these two values is logarithmic and spans about a factor of 100 of strain rate. 
Note that the values reported above as necessary to observe aftershocks ($R/V \gtrsim 100$) correspond to the limit of small velocity in this plot.
A closer examination at the instantaneous 
stress-strain relation reveals that the lower the strain rate, the more pronounced the fluctuations in the
instantaneous stress.

Additional information on the frictional behavior is obtained by studying the system response to abrupt changes of the strain rate. We show in Fig. \ref{ss}B in particular, the stress on a system in which driving is stopped during some time interval (the hold time) and then is re-initiated. First of all, a logarithmic decrease of stress during the hold time is observed. This occurs because the system continues to relax during the hold time and some instability events continue to occur for some time. This is related to our previous statement that aftershocks also occur if driving is stopped after a main shock. Despite the stress reduction during the hold time, a stress peak occurs after re-initiation of sliding.
The height of this peak increases 
logarithmically with the hold time. 
This peak is a consequence of the more stable configuration that the system reached due to relaxation during the hold time. The phenomenon is similar to the one observed in glass forming materials, where it has been explained using the same ideas \cite{jagla}.
These results are in remarkable agreement with those obtained in laboratory measurements \cite{marone,marone2}.

\section{Gutenberg-Richter behavior}

\begin{figure}[t]
\centerline{\includegraphics[width=.4\textwidth]{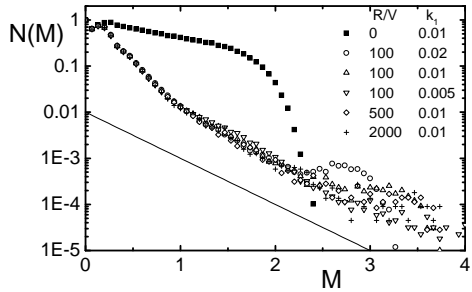}}
\caption{The decaying of number of events with event magnitude. The $R=0$ case is included again for reference.
The thin continuous line shows the case $b=1$, followed rather accurately by actual seismic events. The results in our model in the presence of relaxation show a behavior compatible with the actual seismicity, with a cut-off at large event size that increases upon decreasing the spring constant $k_1$. However, finite size effects seem to be appreciably for the system sizes used. The larger decaying rate observed for $M < 1$ is an spurious effect associated with events comparable in size with our numerical mesh.
}
\label{S6}
\end{figure}

The inclusion of relaxation produces also a change in the decaying exponent $b$ of the GR law.
In Fig. \ref{S6} we see that a power law decaying is maintained in the presence of relaxation, with a $b$ value substantially larger than that corresponding to no relaxation. Once a minimum value of relaxation has been over passed ($R/V \sim 20$), the $b$ decaying exponent is quite insensitive to the precise value of relaxation. There seems to be an excess of events of large magnitude, before the cut off
is reached. The cut off and the peak corresponding to large events are mainly dependent on the value $k_1$ of the spring driving the system. The smaller this value, the larger is the cut-off. It is not clear however if this tendency can be extrapolated to very small values of $k_1$. Unfortunately, to simulate decreasing values of this spring constant requires an increase in system size, and we reach rapidly very time consuming runs. 
The obtained decaying exponent in the presence of relaxation is compatible with the value $b\sim 1$ observed in actual seismicity.

The fact that the $b$ exponent takes a value close to 1 in the presence of relaxation, quite independent of the precise value of the relaxation parameter and other details of the model seems to indicate that relaxation takes the system out of its original universality class with $b\simeq 0.4$,
to a new one with $b\simeq 1.0$. Coincidence of this value with actual ones is another indication that we are capturing essential features of the seismic process with the inclusion of the relaxation mechanism.

\section{Conclusions}

Summarizing, in the present paper we have presented a modeling that combines a spring-block type system in the spirit of the BK model but without a priori velocity weakening friction, with a rather generic implementation of ageing effects within the sliding materials. The motivation for this approach was to introduce, in a spring-block model, a mechanism that generates (and not merely assumes) non trivial frictional effects, which 
can produce realistic temporal and spatial clustering of earthquakes.

Our model allows to obtain a time sequence of events that globally follow the GR law with a $b\simeq 1$ exponent, and at the same time highly non-trivial spatial and temporal correlations compatible, in particular, with the Omori law. In addition we have shown that  frictional properties of the model compare very well with laboratory results. We think this model gives a unified and comprehensive physical picture of all these phenomena.

\section{Acknowledgments}

This research was financially supported by Consejo Nacional de Investigaciones Cient\'ificas y T\'ecnicas (CONICET), Argentina. Partial support from
grants PIP/5596 (CONICET) and PICT 32859/2005 (ANPCyT, Argentina) is also acknowledged.

\end{document}